\documentclass[a4paper]{article}

\usepackage[spanish]{babel}
\usepackage[utf8x]{inputenc}
\usepackage[T1]{fontenc}

\usepackage{enumitem}

\usepackage[a4paper,top=3cm,bottom=2cm,left=3cm,right=3cm,marginparwidth=1.75cm]{geometry}

\usepackage{amsmath}
\usepackage{graphicx}
\usepackage[colorinlistoftodos]{todonotes}
\usepackage[colorlinks=true, allcolors=blue]{hyperref}

\title{Invitación al estudio estadístico del lenguaje\footnote{Este texto es el contenido de una conferencia dictada por el autor en la Universidad de Barcelona en el año 2009. La traducción al castellano es de Abril de 2018. La versión original en catalán fue publicado en Jaume Martí y Marina Salse (coord.) La terminología y la documentación: relaciones y sinergias Barcelona: Instituto de Estudios Catalanes, 2010, p. 47-73 DOI: 10.2436 / 15.2503.02.2 }}
\author{Rogelio Nazar}
\date{}

\begin{document}
\maketitle

\begin{abstract}
El tema de esta presentación es el cruce interdisciplinar entre la lingüística y la estadística. Está dirigida a lingüistas, para los que puede tener un interés teórico, o profesionales que trabajan con la lengua, para los que puede tener un interés práctico. Enfoca el concepto de probabilidad de combinatoria de palabras desde tres perspectivas diferentes: a) los estudios de asociación entre las unidades que se combinan, b) la distribución en el corpus de esta combinación de unidades y, finalmente, c) las formas de medir la similitud entre unidades de acuerdo con sus posibilidades de combinación. Todos estos temas están tratados de una manera estrictamente teórica y van acompañados de ejemplos de aplicación práctica en terminología y en documentación. El objetivo es demostrar que la utilización de herramientas estadísticas en estos campos es un complemento necesario para la intuición de los investigadores. 

\textbf{Palabras clave}: corpus textuales, estadística, lingüística cuantitativa, lenguaje, proprobabilidad combinatoria. 

\textbf{Abstract}: Invitation to the statistical study of language: The topic of this presentation is the interdisciplinary nexus between linguistics and statistics. It targets linguists, for whom it may have a theoretical interest, or professionals that
work with language, for whom it may have a practical interest. It focuses on the concept of
the combinatory probability of words from three different perspectives: a) the studies of association between the units that are combined, b) the distribution of this combination of
units in the corpus, and finally c) the ways of measuring similarity between units according
to the combination possibilities. All these topics are addressed in a strictly theoretical fashion
and are illustrated by examples of practical application in terminology and in documentation. The objective is to demonstrate that the use of statistical tools in these fields is a necessary complement to the researcher's intuition.

\textbf{Keywords}: text corpora, statistics, quantitative linguistics, language, combinatorial probability. 
\end{abstract}

\section{Introducción}

Esta comunicación está dirigida a personas que no tienen conocimientos previos sobre el cruce interdisciplinario entre la lingüística y la estadística. Tiene el doble propósito de ser una aportación desde el punto de vista de la lingüística teórica y, a la vez, ser útil para la documentación y para la terminología. Consecuentemente, incluye ejemplos de cómo este conocimiento teórico se puede aplicar a la solución de problemas prácticos. 

La intención es introducir a la temática, pero también concienciar y aclarar. Concienciar, porque lamentablemente la lingüística cuantitativa no sólo es un área marginal en la lingüística actual sino que, además, muchas veces tanto lingüistas como estadísticos ignoran su existencia. Aclarar, porque la relación entre estadística y lengua no es ninguna novedad ni pertenece al mundo de las ``nuevas tecnologías''. Estamos hablando de una tradición que hace más de sesenta años que difunde conceptos y métodos que no tienen una relación intrínseca con la informática. La utilización de ordenadores es evidentemente necesaria para llevar a cabo estudios en lingüística cuantitativa, pero hablar de estos temas no significa hablar de programas informáticos, porque esto equivale a confundir el fenómeno observado con el instrumento de observación. Ciertamente, los medios son determinantes, ya que, como decía Saussure, el punto de vista define el objeto. Sin embargo, esto no debe llevar al error de reificar las ideas en la forma de un software. En definitiva, lo importante es conocer qué estudios se han hecho o se pueden hacer y tomar conciencia de que esta disciplina no se limita al recuento de veces que dos palabras aparecen juntas en un corpus. 

En cuanto a mi legitimidad como orador, estoy aquí por mi función en el IULA\footnote{Institut Universitari de Lingüística Aplicada (http://www.iula.upf.edu). La relación contractual estaba vigente en el momento de la charla (año 2009).},  consistente en asimilar el conocimiento que ya existe sobre lingüística cuantitativa, aplicar este conocimiento a la solución de problemas prácticos y, a la vez, intentar proponer algún conocimiento nuevo en los foros científicos. No presento nada nuevo en esta comunicación. Haré, en cambio, un recorrido por algunas ideas que he tratado ya en otros trabajos. Es importante advertir que no represento necesariamente la opinión de mis compañeros de trabajo. Me refiero particularmente a un protocolo que incluye un compromiso con la independencia de lengua, es decir averiguar primero hasta qué punto se puede llegar a sacar conclusiones útiles sin introducir conocimiento explícito sobre una lengua en particular. 

Esta comunicación está organizada de la siguiente manera: en la próxima sección 2, analizaremos la confrontación existente entre dos formas muy diferentes de acercarse al estudio de la lengua, ante las que la lingüística se encuentra en una posición ambivalente: el mundo humanístico, por llamarlo de alguna manera aunque parezca ligeramente impreciso, y el mundo científico, en particular el mundo de las «ciencias duras» por oposición a las ciencias sociales, donde el pensamiento cuantitativo es, a veces, todavía visto con sospecha. A continuación, en la sección 3, entraremos en la materia del análisis lingüístico enfocado desde la perspectiva estadística. Analizaremos concretamente el concepto de combinatoria de palabras desde tres perspectivas diferentes: en la subsección 3.1, los estudios de asociación entre las unidades que se combinan; en la subsección 3.2, la manera en que esta combinación de unidades se distribuye en un corpus y las conclusiones que podemos derivar de ello y, finalmente, en la subsección 3.3, las formas de calcular la similitud entre unidades de acuerdo con sus posibilidades de combinación. Como ejemplo, analizaremos el bigrama y estableceremos el significado de esta unidad más allá de su definición formal, para saber en profundidad qué tipo de información codifica. Veremos que, aunque parezca sorprendente, nuestra identidad individual y colectiva está contenida en el bigrama. Como ejemplo de las aplicaciones prácticas, en la sección 4 veremos la clasificación de documentos en distintas variantes (subsecciones 4.1 y 4.2) así como elementos para la caracterización del significado y la desambiguación de terminología. Finalmente, en la subsección 4.3 abordaremos el descubrimiento de neología. 

Existen otras posibilidades de aplicación, entre las que encontramos líneas de investigación en curso, tales como la extracción automática de terminología especializada o la extracción de terminología bilingüe de corpus no paralelos, pero estas líneas, a pesar de su interés, no se tratarán aquí por las limitaciones de espacio.

\section{El choque entre dos culturas}

 Wilhelm Dilthey (1883) advirtió ya las diferencias epistemológicas entre las ciencias naturales, por un lado, y las ciencias sociales y humanidades (o ciencias del espíritu), por otro, continuando una línea de pensamiento que ya había iniciado Kant. Mientras que en las ciencias naturales prevalece un pensamiento mecanicista, con el que se puede predecir la consecuencia de determinados acontecimientos, en las ciencias del espíritu, en cambio, este determinismo no es posible. La respuesta de un ser humano ante un determinado evento es en última instancia imprevisible. Incluso en estas circunstancias, las ciencias del espíritu nos permiten al menos comprender (\textit{verstehen}) las circunstancias históricas e individuales que rodean a lo humano. 

En hito, sin embargo, en la historia de la toma de conciencia de la división de la cultura en el saber científico y el saber humanístico --división que todavía estructura la currícula de la educación secundaria y superior-- sea probablemente una conferencia dada por C.P. Snow (1959), en la que describe la sospecha mutua y la incomprensión existente entre científicos e intelectuales. Aunque pertenezcan a las capas más educadas de la población, los dos colectivos son ignorantes el uno del otro. Si bien después moderó su discurso, en aquella ocasión Snow planteó que la gente que tiene un pensamiento de tipo técnico es en general inculta, y los intelectuales, por su parte, hostiles a este pensamiento técnico, son generalmente incapaces de comprender los conceptos científicos más elementales. 
 
Esta separación de saberes resulta particularmente interesante en el seno de las ciencias sociales, consideradas «ciencias blandas» por oposición al rigor de las ciencias naturales, las «ciencias duras». La inclinación de los científicos sociales por una o por otra rama de pensamiento dependerá de la orientación ideológica personal o de la de cada facultad o departamento, pero entre los intelectuales de las ciencias sociales es común advertir una reticencia \textit{a priori} hacia todo pensamiento de tipo técnico en el estudio de lo que es humano. Esta reticencia está representada en la idea de Castoriadis (1975) sobre el hecho de que con un lenguaje reducido a lo que es instrumental se puede operar y calcular, pero no se puede pensar, una idea con resonancias a la polémica constatación hecha por Heidegger sobre la idea de que «la ciencia no piensa».

En sociología, esta diferencia estuvo claramente representada por la oposición entre el pensamiento crítico y la reflexión filosófica e histórica de la Escuela de Frankfurt frente al hábito de los sociólogos norteamericanos de la Mass Communication Research de promover la aplicación de métodos cuantitativos por encima de la reflexión teórica, enfrentamiento que continuó a pesar de la colaboración entre algunos de los máximos exponentes de ambos bandos, como Theodor Adorno y Paul Lazarsfeld. 

El caso es particularmente interesante en la lingüística, si se quiere, «la más dura de las ciencias blandas». Incluso lingüistas experimentados expresan sorpresa al tomar conciencia de que existe una lingüística cuantitativa. Los que son «de letras» no saben «de números». Mandelbrot (1961) todavía estaba en el momento oportuno para revitalizar la pregunta sobre qué es la lingüística y establecer una diferencia entre gramáticos y lingüistas. En el caso de los primeros, prevalece el conocimiento de una lengua en particular y de lo que puede ser y lo que no puede ser gramaticalmente correcto; mientras que, según este autor, la lingüística pertenece al mundo de las ciencias duras, y, por tanto, para ellos lo importante no son tanto las características particulares de cada lengua, que son de una infinita diversidad, sino las propiedades estructurales del lenguaje (actitud contra la que Saussure seguramente no tendría nada que decir). El estudio de estas propiedades posibilita enunciados científicos con una validez que trasciende el conocimiento que se tenga de una lengua en particular, lo que está de acuerdo con el espíritu científico que es proclive a la generalización, ya que no hay o no debería haber ciencia de lo particular. 

El cruce interdisciplinario, sin embargo, es difícil. Las personas que venimos de ámbitos más cercanos a la lingüística en general estamos poco informados sobre los conceptos matemáticos más elementales y resulta laborioso empezar de cero en el campo, sobre todo para quien no tiene los hábitos de pensamiento de las ciencias duras. Sin embargo, este es, sin dudarlo, un campo de estudio que justifica el desafío. Por ello, por medio de esta presentación, pretendo contagiar el interés y aportar argumentos a la confusión de las barreras entre ciencias duras y blandas, o entre conocimiento científico y conocimiento humanístico en general. Estas barreras ya se confunden y la lingüística no es el único ejemplo. La teoría literaria, mundo humanístico por antonomasia, empieza a sufrir también el asedio de la estadística. Un ejemplo es la aportación que la estadística está haciendo en las disputas sobre la autoría de obras literarias, en casos que incluyen Figuras prominentes como la de Shakespeare (Vickers, 2002). 

\section{La información como probabilidad}

En la línea de Shannon (1948), podemos estimar la cantidad de información como la probabilidad de ocurrencia de un signo en un mensaje, una medida de la cantidad de sorpresa que nos puede provocar un determinado evento. Para explicarlo con palabras sencillas, en determinados contextos sabemos que hay eventos que son más o menos normales y otros que son inesperados. En el lenguaje hay ciertas concatenaciones que son más predecibles que otras. Si cada día, al salir de la trabajo, el jefe dice «hasta mañana» al trabajador, tras una serie de eventos de este tipo el enunciado resulta poco informativo. Pero si un determinado día el texto cambia por «esta empresa ya no requerirá sus servicios», diremos que este segundo enunciado es comparativamente más informativo, es decir que causa mayor sorpresa. Esta sorpresa está directamente relacionada con la probabilidad de aparición de este mensaje (la sorpresa no será tan grande si el empleado está acostumbrado a ser despedido de diferentes trabajos). 

El criterio de la frecuencia como estimación de probabilidad es el mismo que aplicamos cuando nos encontramos en la situación de sacar bolas de una urna. Si suponemos que cada bola tiene la misma probabilidad de ser elegida, y si al sacar las bolas de una en una observamos que las bolas a veces son negros y otras veces son blancas, y tras sacar cien bolas nos damos cuenta de que hemos obtenido noventa cinco bolas negras, esta circunstancia, aunque sea de manera intuitiva, nos hará sospechar que la próxima bola, la 101, tendrá un 95\% de probabilidades de ser negra. Podemos aplicar esta intuición al estudio del lenguaje y adjudicar así un valor que represente la cantidad de información de los signos de acuerdo con su probabilidad de aparición en un mensaje. En la Ecuación 1, la probabilidad de aparición de una determinada palabra $i$ es expresada como $p(i)$, $f(i)$ sería la frecuencia de  $i$ en un determinado corpus y $N$, la cantidad total de palabras de este corpus. 

\begin{equation}
p(i) = f(i) / N 
\label{Eq1}
\end{equation}

En el léxico tenemos palabras que son más o menos informativas. La aparición de palabras como \textit{el, de} o \textit{que} en un texto nos sorprende poco, y por eso decimos que son poco informativas. Si ordenamos todas las palabras de un corpus por frecuencia decreciente, observaremos que la frecuencia de una unidad está en función de su posición en el rango (r); por tanto, se cumple -aproximadamente- la Ecuación [2]:

\begin{equation}
f(x) = 1 / r 
\label{Eq2}
\end{equation}

Si multiplicamos la frecuencia de una unidad por su rango (Ecuación [3]) obtenemos un valor constante.

\begin{equation}
c = f . r 
\label{Eq3}
\end{equation}

La curva de la función [2] representa también la distribución de la renta en las sociedades capitalistas, conocida como la ley de Pareto, por Vilfredo Pareto, quien la describió en 1906. Ordenados de mayor a menor renta, se advierte como son unos pocos los individuos que poseen la mayor parte de la riqueza, mientras que la gran mayoría percibe una mínima parte. Entre los lingüistas, su descubrimiento se atribuye a J. Estoup, quien la describió en 1916, aunque fue divulgada por G. Zipf en 1949. El interés por la ley de Zipf decayó, sin embargo, a partir del estudio de Mandelbrot (1961), quien la reformuló para que se adaptara mejor a los datos observados (Ecuación [4]), particularmente en los rangos más altos y más bajos de la curva.

\begin{equation}
f(x) = P . (r + p)^{-B}
\label{Eq4}
\end{equation}

En la fórmula de Mandelbrot, $f$ es la frecuencia y $r$ el rango, mientras que $P$, $p$ y $B$ son parámetros constantes. Herdan (1964), sin embargo, objeta que estos tres parámetros no son constantes sino que dependen del tamaño del corpus. La consecuencia de esto es que la fórmula no podría ser aplicada a la comparación de muestras de tamaño diferente con el fin de, por ejemplo, comparar la riqueza léxica de las muestras. 

La riqueza del vocabulario está directamente relacionada con la cantidad de información de los signos, lo que determina el grado de dificultad de lectura o densidad de un texto. Esto es lo que Mandelbrot llama la ``temperatura del discurso''. En su caso, planteaba la relación entre la extensión y el vocabulario de un texto, es decir, 
la cantidad de palabras diferentes dividida por la cantidad total de palabras. Pero podemos establecer diferentes medidas de riqueza del vocabulario para un autor o un texto no solamente según esto, sino también poniendo en relación un texto analizado con un conocimiento previo que podamos tener de la lengua en que está escrito. Este conocimiento previo puede tener la forma de un modelo de lengua elaborado sobre la base de un corpus de una extensión de $n$ millones de palabras, un corpus que podríamos llamar corpus de referencia de una lengua, conformado por textos de prensa u otros géneros que pertenezcan a una determinada lengua o variedad dialectal. Mal llamado «corpus de referencia» en realidad, ya que este corpus, por más grande que sea, siempre tendrá un determinado sesgo y no llegará a ser verdaderamente una referencia de la lengua. Utilizado como modelo, sin embargo, nos permitirá conocer la rareza de las palabras que utiliza un texto o un autor, ya que para nosotros representaría un estándar de lo que se puede considerar lengua «normal». 

\subsection{Asociación}

A pesar del interés que pueda tener la asignación individual de información para los signos, es mucho más interesante estimar sus probabilidades de combinatoria. Si los signos se combinaran en el lenguaje de manera aleatoria, sus probabilidades de combinación serían iguales a la multiplicación de sus probabilidades individuales de aparición. La probabilidad de combinación aleatoria de las palabras $i$ y $j$ (Ecuación [5]) define que la probabilidad de aparición conjunta de $i$ y $j$ (expresada aquí como intersección) es igual a la de $i$ multiplicada por la de $j$.

\begin{equation}
p (i \cap j) = p (i) . p (j)
\label{Eq5}
\end{equation}

Existe una abrumadora cantidad y diversidad de medidas para calcular las probabilidades de combinación de las palabras -o eventos, en general- (Muller, 1973; Manning y Schütze, 1999; Evert, 2004; entre otros). En lingüística podemos ver estas medidas aplicadas a la extracción de terminología especializada polilexemática o al estudio de las colocaciones, todo un capítulo en el estudio del lenguaje. Las combinaciones de palabras no son dadas solamente por la gramática, y esto tiene indudablemente su correlato en las frecuencias de coocurrencia. Por ejemplo, en inglés, se dice \textit{strong coffee}, pero no \textit{powerful coffee}. Sin embargo, podemos hablar de una \textit{powerful computer}, pero no de una \textit{strong computer}\footnote{Este último ejemplo es interesante, porque actualmente ambas secuencias de palabras tienen prácticamente la misma frecuencia en Google; algo que puede engañar al usuario desprevenido porque la segunda forma, \textit{strong computer}, aparece siempre formando parte de estructuras más grandes como \textit{strong computer password}. Es decir, el núcleo del que depende \textit{strong} no es en este caso \textit{computer} sino \textit{password}, o \textit{skills}, o \textit{science, background}, etcétera.}. 

En cada lengua, e incluso en cada
dominio de especialidad, existen ciertas preferencias en las combinaciones de palabras en las diversas categorías gramaticales (verbo-nombre; adjetivo-nombre; nombre-nombre, etc.). Por una razón pragmática, las cosas suelen decirse de una determinada manera, y si bien la gramática nos permitiría formular el texto de otra, haciéndolo así correríamos el riesgo de confundir al receptor si ya existe, en esta lengua, dominio o registro, una manera típica o idiosincrática de decir lo que queremos decir. 

Las estadísticas de asociación nos pueden informar sobre la manera típica en la que se combinan las palabras de una lengua porque responden a la pregunta sobre cuál es la probabilidad de que dos eventos ocurran juntos en una misma situación, o más precisamente, si la frecuencia de aparición de dos eventos en una misma situación se puede atribuir al azar. 
Un evento puede ser la aparición de una palabra y la situación puede ser un texto, un párrafo, una oración, una «ventana» de $n$ palabras, etc. También se puede tratar de la aparición de las palabras de forma concantenada, o no. Si se trata de una secuencia de dos palabras podemos hablar de un bigrama, de un trigrama en el caso de tres unidades o de un $n$-grama para $n$ unidades. Pero hay que tener en cuenta que un $n$-grama podría ser definido de otra manera, como una secuencia de letras o de categorías gramaticales. 

La coocurrencia, además, puede ser definida de una manera distinta a la secuencial. Podemos definir coocurrencia como la aparición de las dos palabras en una ventana de contexto sin importarnos el orden en que aparecen. Las Figuras 1 y 2 muestran, por ejemplo, un criterio de coocurrencia que consiste en comprobar cuántas veces aparecen las palabras --a diferentes distancias y en diferente orden-- en una ventana de contexto de veinte palabras.  En ambos casos, estamos analizando las palabras que coocurren con la forma inglesa \textit{platypus} (ornitorrinco) en un corpus descargado de Internet.

\begin{figure}
\centering
\includegraphics[width=0.7\textwidth]{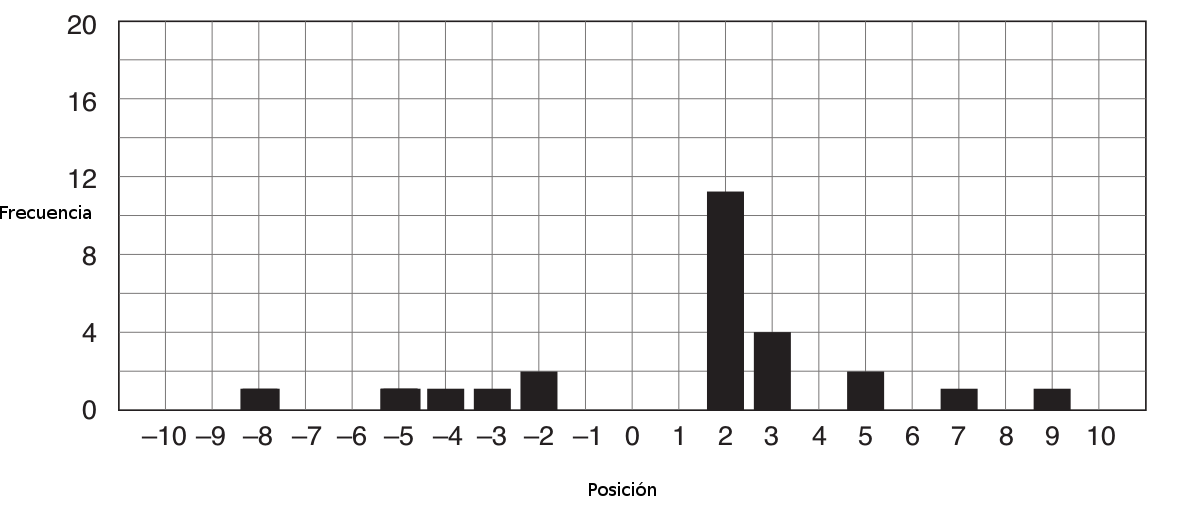}
\caption{\label{fig1}Histograma que caracteriza la coocurrencia de la forma platypus (`ornitorrinco', en inglés) y la forma anatinus (parte de su denominación científica). Ejemplo 1: \textit{...the platypus ornithorhynchus anatinus is a semiaquatic mammal endemic
to eastern Australia, including...}}
\end{figure}

En la Figura 1, observamos que las ocurrencias de la forma \textit{anatinus}, una de las palabras con las que está asociada, se reparten a izquierda y derecha de \textit{platypus}. Comprobamos aquí que las ocurrencias de \textit{anatinus} se concentran en la posición +2, es decir que la mayoría de las veces la forma \textit{anatinus} aparece dos posiciones tras la forma \textit{platypus}, como en el ejemplo 1. En la Figura 2, observamos que lo mismo ocurre con la forma \textit{has}, aunque ahora la forma se concentra en la posición +1, tal como ocurre en el ejemplo [2].

\begin{figure}
\centering
\includegraphics[width=0.7\textwidth]{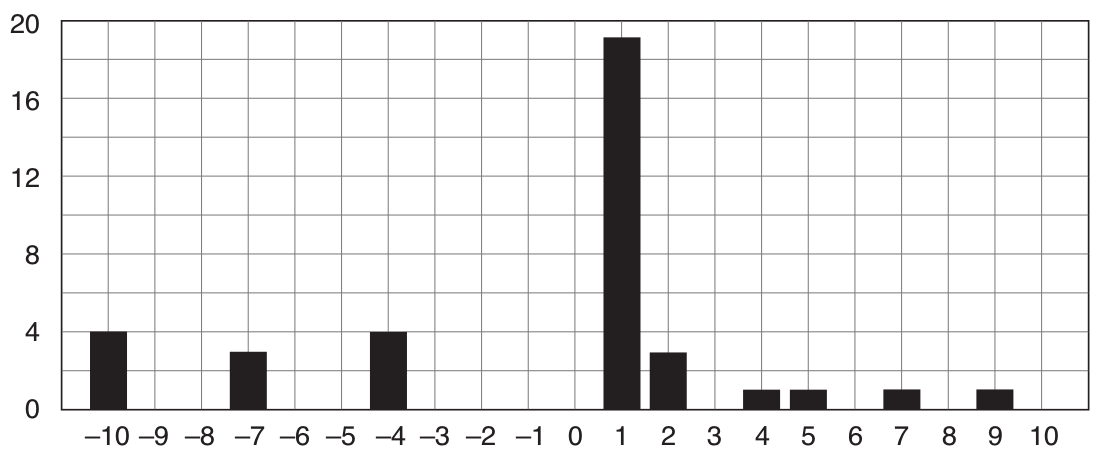}
\caption{\label{fig2}Histograma por las formas platypus y \textit{has} ( 'tiene'). Ejemplo 2: \textit{...the platypus has four legs which extend horizontally from its body...}}
\end{figure}

En lingüística de corpus es habitual utilizar medidas de asociación, pero no tanto para falsar una hipótesis nula según la cual los elementos que estamos estudiando se combinarían por azar, sino más bien para ordenar combinaciones de elementos a partir de la ponderación que obtienen como consecuencia de la aplicación de estas medidas. 

Podemos establecer diferentes tipos de medidas de asociación en función de la simetría o asimetría que presentan. Entre las medidas de asociación simétricas encontramos el concepto de información mutua (Ecuación [6]), derivado de la teoría de la información de Shannon (1948). Representa la cantidad de información que nos da la ocurrencia del evento $i$ sobre la ocurrencia del evento $j$ (Church y Hanks, 1991; Manning y Schütze, 1999). Con esta Ecuación medimos, en bits, qué tan previsible es un acontecimiento $i$ al pasar uno $j$. Es decir, cuánta sorpresa nos causa $i$ una vez que ha aparecido $j$. En un caso extremo, una alta información mutua sería que $i$ sólo ocurre cuando ha pasado $j$, y en el extremo opuesto, que si pasa $i$ puede pasar $j$ o cualquier otro acontecimiento. Es simétrica por definición, ya que otorga un mismo valor a $i$ dada $j$, que a $j$ dada $i$. Esta medida no es aplicable a eventos que tienen poca frecuencia, ya que atribuiría una alta asociación a los que aparecen de manera conjunta por simple azar.

\begin{equation}
MI(i, j) = \log_2 \frac{p(i,j)} {  p(i) p(j) } 
\label{Eq6}
\end{equation}

En tanto, entre las medidas de asociación asimétricas encontramos la probabilidad condicional (Ecuación 7). Es una medida asimétrica porque puede no ser igual la probabilidad $i$ dada $j$, que la probabilidad de $j$ dada $i$. Por ejemplo, si $j$ es la palabra \textit{augurio} e $i$ es \textit{mal} (o \textit{buen}), la palabra \textit{augurio} predice \textit{mal}, pero \textit{mal} no predice en absoluto a \textit{augurio}.

\begin{equation}
p(i | j) = \frac{p(i \cap j)} { p(j) } 
\label{Eq7}
\end{equation}

Hasta ahora hemos visto ejemplos con bigramas, es decir, secuencias de dos palabras. Si estamos estimando la probabilidad de aparición de un bigrama, podríamos también volver a la Ecuación [1] y definir la probabilidad como la frecuencia de aparición dividido por la cantidad total de bigramas que hemos observado en el corpus. Veremos en la sección 4 que es posible, estudiando sólo las frecuencias de aparición de los bigramas, reconocer la escritura de autores individuales. Esto es posible porque el lenguaje es un sistema de opciones y elecciones. El lenguaje ofrece al hablante o autor diferentes posibilidades de combinatoria, y este, con las sucesivas elecciones, se va construyendo a sí mismo. Entonces empiezan a producirse combinaciones que son recurrentes o típicas de un autor en comparación con otros. Pero no hablamos sólo de autores, porque también las variantes dialectales de los diferentes colectivos o naciones tienen una determinada manera de combinar las palabras y conforman patrones que el ordenador puede reconocer mediante la aplicación de un sencillo cálculo estadístico. Estos patrones, no hace falta decirlo, son completamente imperceptibles para el ojo humano, y quien los produce no es conciente de que lo hace.

\subsection{Distribución}

La sección anterior ofrece una visión del corpus como un espacio continuo donde se puede dar la coocurrencia de eventos-palabras, valiéndose de la noción de ventana de contexto para definir cuándo dos palabras aparecen juntas. Esta sección, en cambio, ofrece una perspectiva diferente del corpus, ya que lo presenta dividido según un criterio determinado. 

En primer lugar, comentaremos algunos ejemplos de cómo podemos estudiar o visualizar la distribución de unidades o de combinación naciones de unidades en corpus divididos de manera diferente. Finalmente, estudiaremos la manera de ordenar las unidades de un corpus a partir del comportamiento que tiene su curva de distribución.

El primer ejemplo es el análisis de la distribución de términos en un documento concreto. De acuerdo con finalidades diversas, ya sea el análisis del discurso en el plano teórico o la elaboración de sistemas de indexación para la recuperación de información, podemos tener interés en averiguar cómo se distribuyen las ocurrencias de determinados términos en la obra de un autor. Es posible que existan términos clave en ciertas obras que se distribuyan de una manera recurrente a lo largo del texto. También puede ocurrir que algunos términos se concentren en determinados capítulos de la obra. Puede que se encuentren en la introducción, por ejemplo, ya que su función es introducir al lector en los conceptos que luego presentará el texto, asociados a los conocimientos que se supone tiene el lector. Pero es posible también que estos términos introductorios no sean fundamentales en la obra. La Figura 3, por ejemplo, muestra que tres términos clave en una versión en inglés de la Crítica de la Razón Pura, de Kant: \textit{concepts}, \textit{empirical} e \textit{intuition} se distribuyen de manera regular en la obra, si bien \textit{intuition} se concentra en el capítulo dedicado a la estética. 

También es posible que una gran cantidad de palabras se distribuya de manera regular a lo largo de la obra; pero no porque estas sean importantes en el contenido, sino porque forman parte del sistema de la lengua. Por ello, para los estudios de distribución de una obra concreta, hay que tener en cuenta también la distribución de las unidades en un corpus. 

La Figura 4 muestra un ejemplo de distribución de unidades, esta vez en un corpus diacrónico. Se trata de las frecuencias de las palabras en los archivos del diario El País\footnote{Véase http://www.elpais.es.}. Cada una de las divisiones en el eje horizontal representa todas las ediciones de un mismo año. El eje vertical representa la frecuencia relativa de una palabra determinada o de una combinación de palabras en cada año. Podemos observar que, mientras que algunas palabras tienen un uso continuo a lo largo del tiempo, ya que son palabras del vocabulario central de la lengua (Figura 4), otras unidades tienen un uso que fluctúa, ya que hacen referencia a conceptos extralingüísticos que tienen diferente vigencia en función de la agenda temática de los medios de comunicación (Figura 5).

\begin{figure}
\centering
\includegraphics[width=0.7\textwidth]{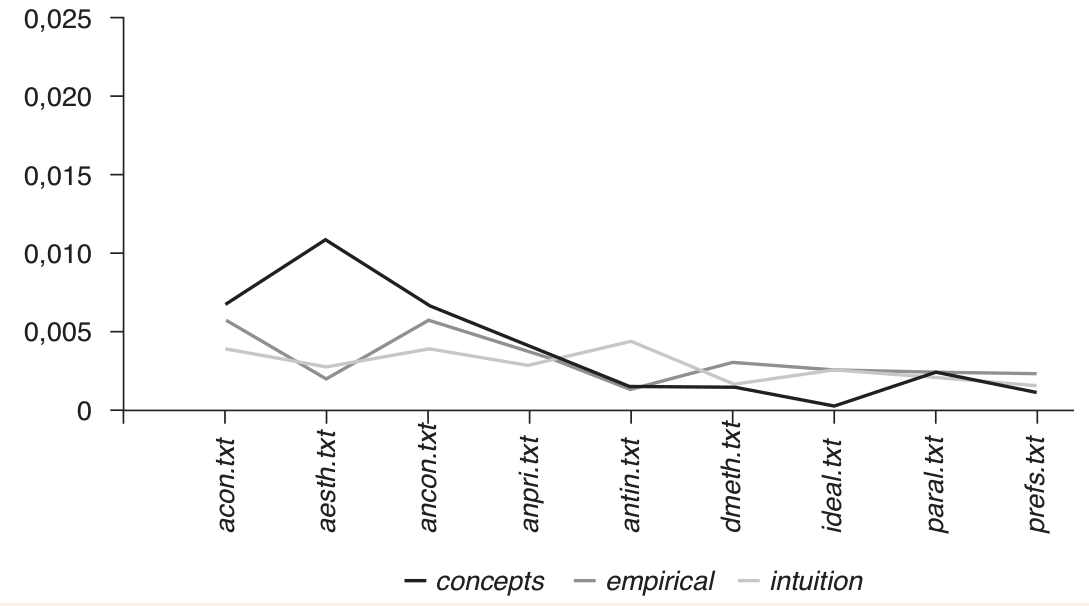}
\caption{\label{fig3}Distribución de las formas \textit{concepts}, \textit{empirical} e \textit{intuition} a lo largo de los diferentes capítulos de una versión en inglés de la ``Crítica de la razón pura'', de Kant. El eje horizontal representa los diferentes capítulos y el eje vertical la frecuencia relativa.}
\end{figure}

\begin{figure}
\centering
\includegraphics[width=0.7\textwidth]{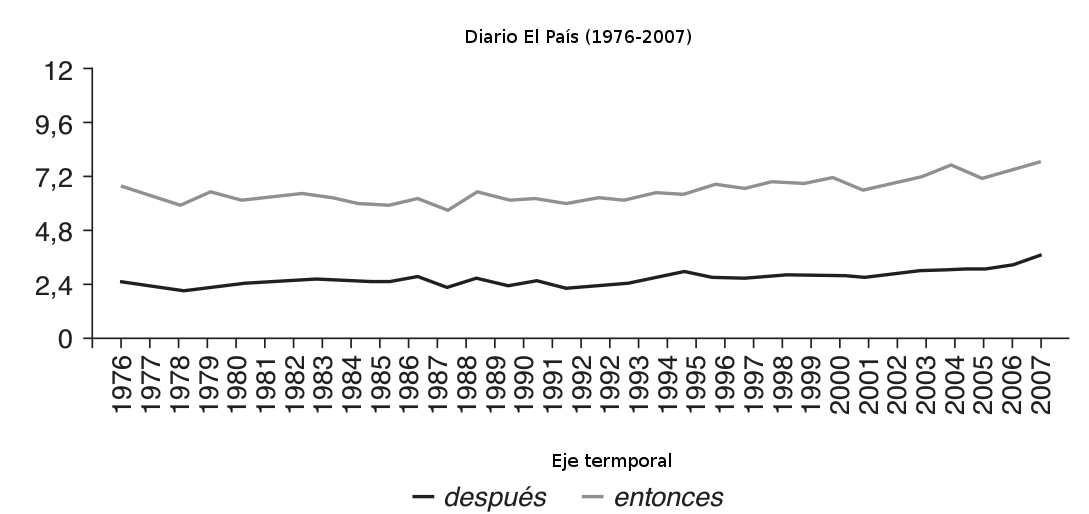}
\caption{\label{fig4}Distribución de las formas \textit{después} y \textit{entonces}, dos palabras del vocabulario central de la lengua castellana, en los archivos del diario El País en el período 1.976-2.007. El eje horizontal representa el tiempo y el eje vertical, la frecuencia relativa.}
\end{figure}

\begin{figure}
\centering
\includegraphics[width=0.7\textwidth]{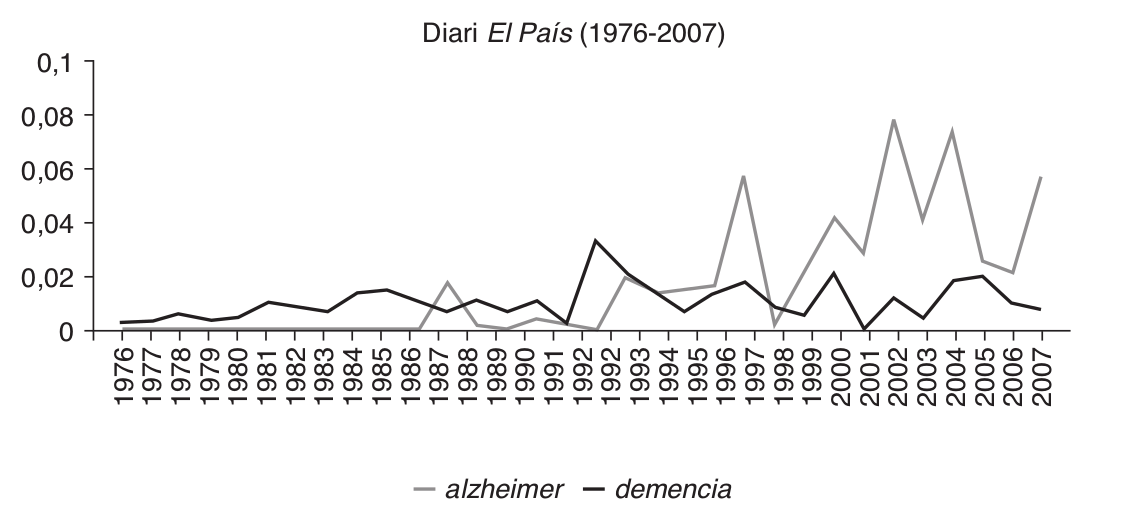}
\caption{\label{fig5}Distribución de las formas \textit{demencia} y \textit{Alzheimer}, en el mismo corpus, dos unidades que hacen referencia a conocimiento extralingüístico.}
\end{figure}

Un caso diferente es el de dos unidades que, si bien también están implantadas, presentan oscilaciones debido a la evolución del sistema semántico de la lengua. Lo ejemplificamos en la Figura 6, con las unidades \textit{hombre} y \textit{mujer}, que representan el desarrollo ideológico de una sociedad que toma conciencia del lenguaje sexista. Así, vemos que mientras en 1976 la palabra \textit{hombre} era mucho más frecuente que la palabra \textit{mujer}, esta diferencia se va revirtiendo con el tiempo hasta alcanzar la misma frecuencia de uso en 2007. 

Basándonos en el comportamiento de las curvas de distribución de frecuencias de las unidades en estos corpus divididos, hay varios coeficientes que nos interesan para diferentes fines. En algunos casos, nos interesarán las unidades o combinaciones de unidades que tengan una frecuencia de uso ascendente, como en el caso de la extracción de neología (subsección 4.3). Pero en otros casos nos interesará saber cuál es el vocabulario consolidado de una lengua, por contraste con las unidades referenciales, es decir, aquellas que hacen referencia a conocimiento extralingüístico. En este caso, nos interesan aquellas unidades que tengan las curvas más horizontales. En el caso opuesto, podemos caracterizar la irregularidad de una distribución mediante la fórmula [8] (Nazar, 2008) que mide la dispersión $D$ de una unidad $t$ mediante la multiplicación del valor máximo de frecuencia de $t$ o $max f ( t )$, que sería la frecuencia de $t$ en la partición donde es más frecuente, multiplicada por $Cr ( t )$, que sería la cantidad de particiones en las que $t$ tiene frecuencia 0 o una frecuencia inferior a un determinado parámetro $k$.

\begin{equation}
D(t) = \max_{f(t)} . Cr(t) 
\label{Eq8}
\end{equation}

\begin{figure}
\centering
\includegraphics[width=0.7\textwidth]{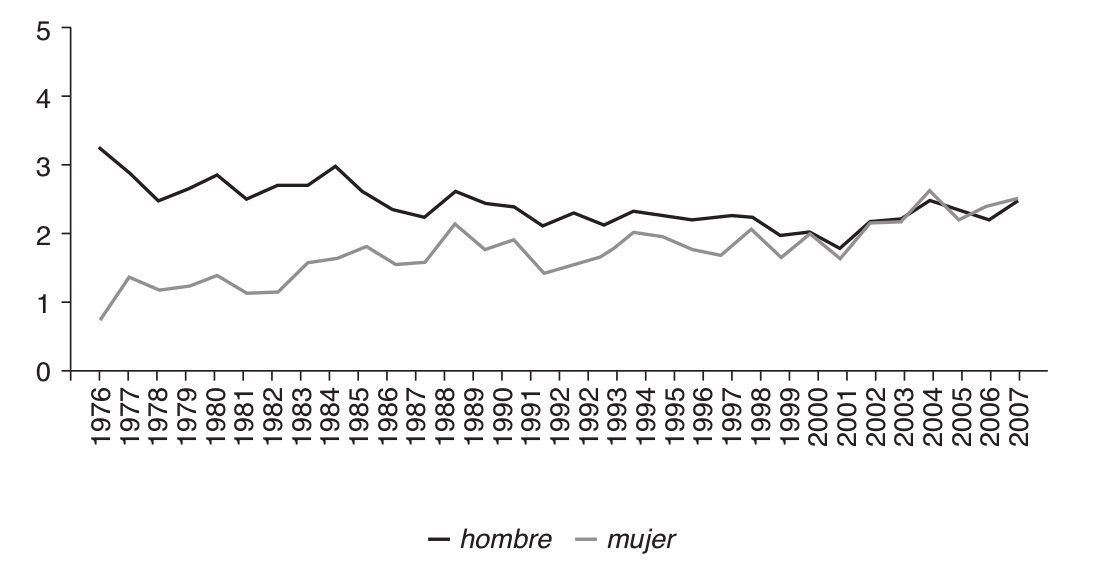}
\caption{\label{fig6}Distribución de las formas \textit{hombre} y \textit{mujer} en el mismo corpus.}
\end{figure}

\subsection{Similitud}

En esta sección tratamos el concepto de similitud desde un espectro amplio. Podríamos hablar exclusivamente de similitud entre entidades lingüísticas, pero hay saber que también es posible calcular la similitud entre diferentes objetos complejos si somos capaces de codificarlos como vectores. 

Podemos agrupar diferentes objetos según la similitud que tengan, definida de acuerdo con los atributos que comparten. Estos 
atributos estarán definidos para cada objeto en forma de vector. Un vector puede representar varias cosas: un documento, el haz 
de coocurrencia de un término, los predicados con los que suele aparecer un nombre, etc. La cantidad de valores de un vector es 
lo que determina su dimensionalidad, $n$, donde los $x$ son los componentes (Ecuación [9]).

\begin{equation}
\vec{x}= ( x_1, x_2, x_3, ..., x_n )
\label{Eq9}
\end{equation}

Un vector se intuye con facilidad como una fila de una matriz. La Tabla 1 muestra, por ejemplo, una matriz de documentos por 
términos, mientras que la Tabla 2 muestra una matriz de términos por términos. 

\begin{table}
\centering
\begin{tabular}{c|c|c|c|c}
\hline
& $Term_1$ & $Term_2$ & $Term_3$ & ... \\
\hline
$Doc_1$ & 1 & 0 & 1 \\
$Doc_2$ & 0 & 1 & 1 \\
$Doc_3$ & 0 & 1 & 0 \\
\hline
... & ... & ...  & ... & ... \\
\hline
\end{tabular}
\caption{\label{docxterm}Matriz de documento por término.}
\end{table}

\begin{table}
\centering
\begin{tabular}{c|c|c|c|c}
\hline
& $Term_1$ & $Term_2$ & $Term_3$ & ... \\
\hline
$Term_1$ & 1 & 0 & 1 \\
$Term_2$ & -- & 0 & 1 \\
$Term_3$ & -- & -- & 0 \\
\hline
... & ... & ...  & ... & ... \\
\hline
\end{tabular}
\caption{\label{termxterm}Matriz de término por término }
\end{table}

Si los objetos que estamos comparando fueran términos y los componentes de sus vectores representaran los $n$-gramas de 
letras que los conforman, entonces podríamos utilizar las medidas de similitud entre cadenas de caracteres para tener, entre 
otras cosas, una forma de pseudolematitzación en el trabajo con textos no etiquetados, ya que esta metodología sería capaz de 
detectar la similitud que existe entre cadenas como \textit{enfermedad} y \textit{enfermedades}; o bien la identificación de 
variantes terminológicas, como en el caso de \textit{superficie pulmonar} y \textit{superficie de los pulmones}. 

Con medidas 
de similitud como estas podemos elaborar, por ejemplo, un programa que, a partir de un término de entrada, indique una lista de 
términos en un corpus que presenten una similitud morfológica. Lo mismo puede hacerse con documentos: a partir de un 
documento determinado, el programa ordenará el resto de los documentos del corpus de acuerdo con la similitud que tengan. 
Las posibilidades, sin embargo, son aún mayores. Por ejemplo, con una colaboradora, Vanesa Vidal, hicimos un experimento en el que 
comparamos diferentes verbos especializados en función de los nombres con los que estos verbos suelen aparecer.

\begin{table}
\centering
\begin{tabular}{c|c|c|c|c|c}
\hline
& $Nom_1$ & $Nom_2$ & $Nom_3$ & $Nom_4$ &... \\
\hline
$Verb_1$ & 0 & 0 & 0 & 1 & ...\\
$Verb_2$ & 1 & 0 & 0 & 0 & ...\\
$Verb_3$ & 0 & 0 & 1 & 0 & ...\\
\hline
... & ... & ...  & ... & ... & ...\\
\hline
\end{tabular}
\caption{\label{verbxnom}Matriz de verbos por nombres.}
\end{table}

La tabla 3 
muestra un fragmento de una matriz que tiene cientos de filas y columnas que cruzan la información de coocurrencia de verbos 
(filas) y nombres (columnas) en un corpus de genómica. Es una matriz binaria, ya que codifica, en cada celda, la aparición o la no 
aparición de las combinaciones verbonominales. La comparación automática de todos los verbos entre sí da una lista de los 
grupos de verbos más similares, es decir, aquellos que se relacionan con el mismo o casi con el mismo grupo de nombres. De este 
modo, podremos ver que, sin tener en cuenta ningún tipo de información sobre la similitud morfológica y ortográfica, 
encontramos que, en castellano, en el ámbito de la genómica, los verbos \textit{enrollar} y \textit{desenrrollar} son muy 
similares porque aparecen junto a los nombres \textit{hélice, cadena, adn, hebra}, etc.; así como los verbos \textit{beber}, 
\textit{ingerir} y \textit{reabsorber} parecen porque comparten los nombres \textit{agua}, \textit{cantidad}, \textit{cola}, \textit{célula} y \textit{glucosa}, entre otros. 

Diferentes autores han adoptado estrategias más o
menos similares, no ya en el estudio de combinaciones verbonominales, sino para el descubrimiento de sinónimos, 
cuasisinónimos o equivalentes en diferentes lenguas que ponen en relación elementos que comparten los mismos vecinos (Nazar, 
2010\footnote{La investigación estaba en curso en el momento de la conferencia.}). Entre otras medidas de similitud, la medida Dice es apropiada para la comparación de vectores con valores binarios. Lo 
que esta medida hace es contar la cantidad de dimensiones en que en dos vectores el valor es 1 en relación con la cantidad de valores. Si X e Y son dos vectores, la 
medida queda expresada en la Ecuación [10]. | X | es el conjunto cardinal de X, es decir, la cantidad de componentes. Se 
multiplica por 2 para tener un escala que va de 0,0 a 1,0, que sería la similitud total.

\begin{equation}
Dice(X,Y) = \frac{ 2 |X \cap Y| } { |X| + |Y|} 
\label{Dice}
\end{equation}

La medida Jaccard (Ecuación [11]) es similar a la anterior, pero introduce una normalización que consiste en la división por la 
cantidad de dimensiones de los vectores, es decir que introduce una penalización cuando hay pocas dimensiones compartidas en 
proporción a la cantidad total de dimensiones.  

\begin{equation}
Jaccard(X,Y) = \frac{| X \cap Y | } { |X \cup Y |} 
\label{Jaccard}
\end{equation}





\section{Aplicaciones prácticas}

Si bien la sección anterior ya sugiere algunos ejemplos de aplicación práctica, en esta sección presentaremos un espectro de 
aplicación más amplio. Analizaremos la aplicación de medidas de similitud y coocurrencia en el ámbito de la clasificación 
automática de documentos en las dos modalidades en que esta práctica existe actualmente: la clasificación con aprendizaje
supervisado y no supervisado. Finalmente, comentaremos brevemente la aplicación de medidas de distribución aplicadas al 
descubrimiento de neología. La falta de espacio nos obligará a dejar temas que habría sido muy interesante comentar, 
como por ejemplo la aplicación de metodologías estadísticas para la extracción de terminología especializada, así como las 
metodologías para la extracción de terminología bilingüe a partir corpus no paralelos, que son líneas de invsestigación en curso.

\subsection{Clasificación de documentos} 

Como es sabido, los algoritmos de clasificación automática de documentos se dividen en supervisados y no supervisados 
(Manning y Schütze, 1999; Sebastiani, 2002).
En ambos casos estamos agrupando objetos (documentos, en este contexto), pero la diferencia es que, en el primero, un 
algoritmo de clasificación tiene un conocimiento previo sobre los objetos que ha de clasificar, ya que ha pasado por un proceso 
de «entrenamiento», en el que un usuario le ha enseñado ejemplos de objetos clasificados según un criterio cualquiera. En el 
segundo caso, en cambio, la tarea de clasificación se hace sin este conocimiento, es decir que el algoritmo no sabrá cuántas ni 
cuáles son las categorías según las cuales los objetos deben ser agrupados, y por tanto la clasificación será una propiedad 
emergente a partir de las similitudes que tienen los objetos. 

\subsubsection{Clasificación con aprendizaje supervisado}

En el año 2004 me 
vinculé a dos grupos de investigación que estaban trabajando en áreas que en principio pueden parecer disímiles. Uno de los 
grupos estaba trabajando en la atribución de autoría con el propósito de aplicarla a la lingüística forense. El otro grupo, más 
vinculado a la terminología, tenía interés en encontrar una manera sistemática de clasificar un documento, tanto según la 
temática como según el grado de especialidad. La filosofía de trabajo en ambos grupos era la misma: diseñar estrategias 
fundamentadas en el conocimiento lingüístico, entendida como el examen manual de la casuística y la identificación, de acuerdo 
con la intuición del investigador, de aquellos rasgos que podrían ser discriminantes de las diferentes categorías.  En ambos 
casos se trata de un trabajo de enorme complejidad y arraigado en el conocimiento que el investigador tiene de la lengua 
particular en la que está escrito el texto. 

En el caso de la lingüística forense, los rasgos pueden ser, por citar algunos ejemplos, 
giros idiosincrásicos que puedan delatar una pertenencia a una zona geográfica o a una condición social, o bien particularidades 
como los errores de ortografía o gramática que tengan en común los textos de autoría disputada con aquellos textos de autoría 
indubitada (véase Turell, 2005, para una introducción). En el caso de la clasificación de documentos por tema o por grado de 
especialidad, la estrategia consistía en encontrar rasgos lingüísticos de un dominio temático (la densidad de terminología 
especializada en el texto, por ejemplo) o bien otros rasgos morfológicos y léxicos que pueden ser característicos de la literatura 
especializada (Cabré et al., 2009). En este contexto surgió el software Poppins\footnote{El programa Poppins puede ser ejecutado a través de Internet en la dirección http://
www.poppinsweb.com} (Figura 7).

\begin{figure}
\centering
\includegraphics[width=0.7\textwidth]{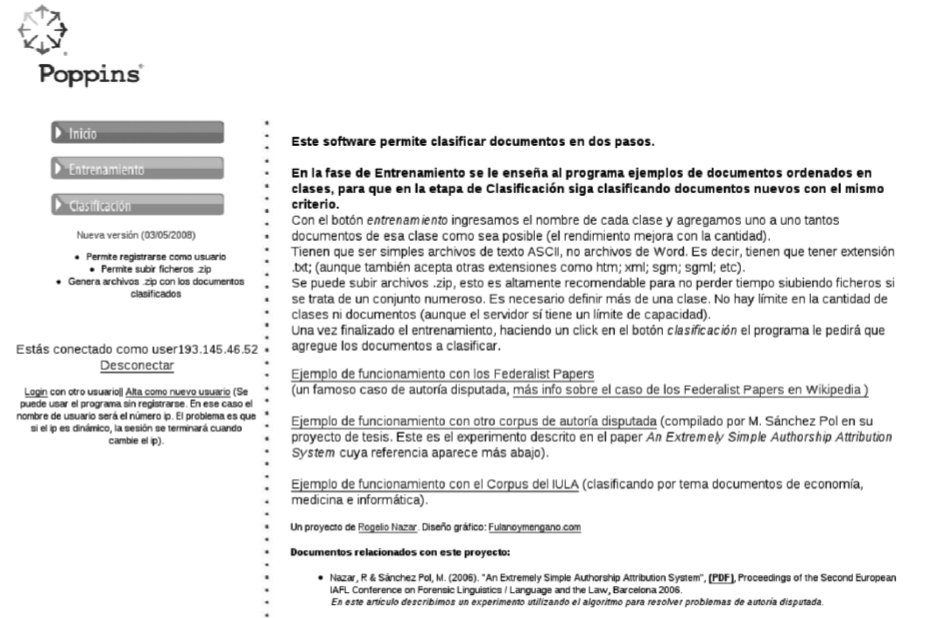}
\caption{\label{fig7}Captura de pantalla de la interfaz web del software Poppins}
\end{figure}

Este programa representa una solución de 
clasificación diferente, ya que se puede aplicar tanto a los problemas de atribución de autoría como la clasificación por tema, por 
grado de especialidad y hasta para otros problemas de clasificación en los que el algoritmo sea entrenado, y ello con 
independencia de la lengua de los documentos, del dominio temático o del criterio de clasificación. Como decíamos antes para el 
caso de los algoritmos supervisados, la lógica de este programa incluye dos fases principales. En la primera, la fase de 
entrenamiento, un usuario «presenta» al programa ejemplos de documentos ordenados en clases. Una vez terminada esta 
etapa, la etapa de clasificación consiste en, partiendo de un nuevo conjunto de documentos, ordenarlos basándose en la 
clasificación que ha aprendido durante la fase de entrenamiento. 

El modo de funcionamiento es bastante básico porque los textos que 
son clasificados no son sometidos a ningún tipo de procesamiento. La única operación que se hace es calcular las frecuencias de 
aparición de los diferentes bigramas de palabras del corpus. Así, cada clase de entrenamiento se convierte en un vector que tiene 
por atributos los bigramas y por valor la frecuencia de aparición. De esta manera, a partir de un nuevo documento, lo que 
hacemos es computar una medida de similitud que consiste en sumar las frecuencias de los bigramas que tienen en común 
el documento a clasificar y cada una de las clases. La comparación que obtiene como resultado la suma mayor es la clase 
elegida para este documento. 

Con otra colaboradora, Marta Sánchez Pol (Nazar y Sánchez Polo, 2006), descubrimos que con este programa
podíamos determinar correctamente la autoría de un texto con una probabilidad del 90\%. La interfaz del programa muestra 
experimentos con otros casos, como el de los Federalist Papers, un famoso caso de autoría disputada, y atribuye los textos de 
autoría disputada a James Madison (Figura 8), lo que coincide con otros estudios llevados a cabo sobre el mismo caso (Mostellaria y Wallace, 
1984). 

En cuanto a la clasificación por temática y por grado de especialidad, experimentos de clasificación de documentos del 
Corpus Técnico del IULA (Vivaldi, 2009) demostraron niveles de precisión similares. El experimento aún se puede repetir de diversas maneras, 
mediante la clasificación los documentos por lengua, por variante dialectal, por género o por otros criterios.

\begin{figure}
\centering
\includegraphics[width=0.7\textwidth]{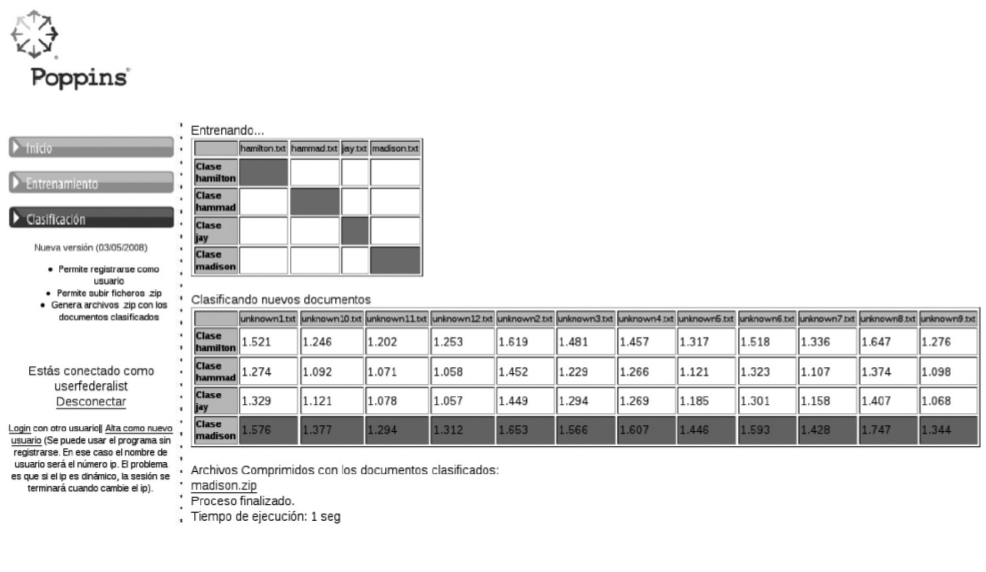}
\caption{\label{fig8}Resultados de la clasificación de documentos con el software Poppins}
\end{figure}

\subsubsection{Clasificación con aprendizaje no supervisado}

Como hemos dicho en la introducción de esta sección, la clasificación con aprendizaje no 
supervisado es el escenario en el que el algoritmo no ha pasado por una etapa de entrenamiento y, por tanto, no sabe cuáles 
ni cuántas son las categorías en las que deben ser clasificados los documentos. Si en el caso anterior relacionábamos la clasificación de documentos con aplicaciones concretas como la atribución de autoría, en este caso la clasificación de documentos con 
aprendizaje no supervisado se peude relacionar con la desambiguación de terminología. Esto es así porque planteamos la 
clasificación como un problema de desambiguación. En este experimento, reunimos una colección de documentos en que 
aparece una forma ambigua, por ejemplo mediante la descarga de documentos de Internet, y los clasificamos a partir de los 
diferentes sentidos que puede mostrar esta forma dentro de la colección. 

Esta clasificación  se lleva a cabo por medio de los grafos de coocurrencia léxica. Tomemos, en primer lugar, un ejemplo con 
una forma ambigua como \textit{ratón} en castellano, que, en el Corpus Técnico del IULA --que contiene documentos 
sobre informática y sobre genómica-- puede ser utilizada para hacer referencia al dispositivo periférico del ordenador o bien al 
animal de laboratorio. En los grafos de coocurrencia hay un nodo principal, situado en la zona superior central, que corresponde 
a la unidad que estamos analizando: \textit{ratón} en este caso. De este nodo dependen todos los demás. Cada nodo representa una 
palabra o una combinación de palabras, y las conexiones entre nodos expresan que las palabras que los nodos representan 
aparecen juntas en los mismos contextos donde aparece la unidad analizada. En la Figura 9 se aprecia la 
existencia de dos regiones en el grafo, una en la derecha y otra a la izquierda. Estas dos regiones --atractores o clústers de nodos-- 
se corresponden con cada uno de los sentidos que la forma presenta. En un caso, las unidades con las que 
aparecerá ratón serán \textit{cromosoma, mamífero, rata, genoma, laboratorio, bacteria}, entre otros; mientras que, en el otro caso, 
las unidades que se relacionan con ratón son \textit{usuario, pantalla, teclado, clic}, etcétera. 

En la tesis (Nazar, 2010) presento, entre 
otras cosas, la aplicación de este método para la desambiguación de siglas, ya que estas son formas ambiguas por naturaleza. Así, ante de una 
colección de documentos descargada de Internet con una forma ambigua como NLP, por ejemplo, un programa informático es 
capaz de obtener dos clústers que representan los dos sentidos de esta palabra: por un lado, documentos referidos a la forma 
expandida \textit{natural language processing} y por otro, documentos sobre \textit{neurolinguistic programming}. En el primer caso, NLP 
se relaciona con unidades \textit{comon knowledge representation, language technology, functional grammar, machine translation, statistical NLP, computational lingüísticos}, entre otros; mientras que el segundo clúster incluye unidades como \textit{practitioner 
training, practitioner NLP, gestalt therapy, John Grinder, Richard Bandler, Robert Dilts}, etcétera.

\subsection{Descubrimiento de neología} 

En esta sección analizaremos la aplicación de algunas de las medidas de distribución que hemos visto en la sección 3.2, 
con el propósito concreto de hacer un experimento de extracción automática de neología. Los resultados de la aplicación de estas técnicas para la extracción de neología, así como de las técnicas de desambiguación automática presentada en el punto 
anterior (4.1.2) fueron presentadas en un trabajo previo (Nazar y Vidal, 2008).  


\begin{figure}
\centering
\includegraphics[width=1\textwidth]{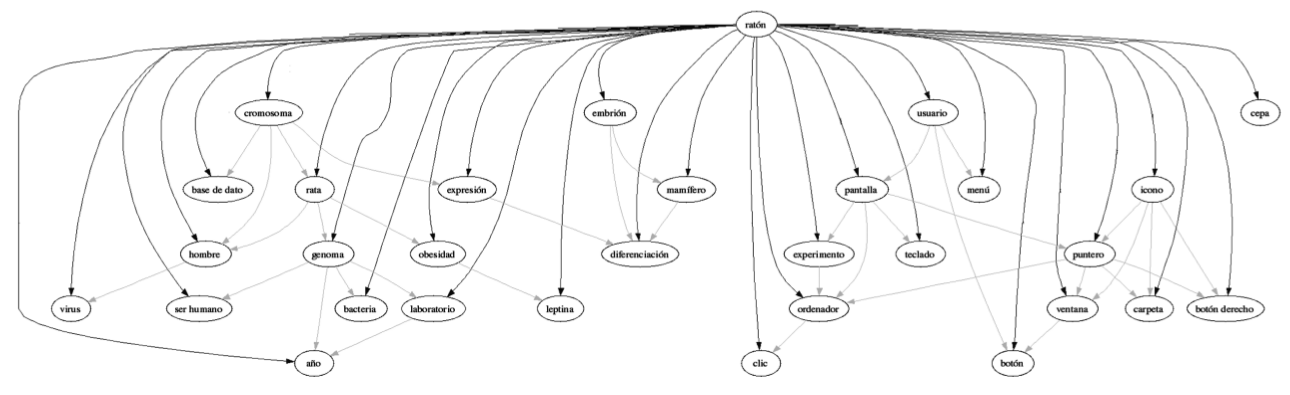}
\caption{\label{fig9}Grafo de coocucurrencia de la forma polisémica ratón, con un significado que hace referencia al dispositivo periférico del ordenador en los documentos de informática y el otro utilizado en los documentos de genómica para hacer referencia al animal, frecuentemente utilizado en laboratorios.}
\end{figure}

Las Figuras 10 y 11 ofrecen gráficas que ya nos son familiares, porque hemos visto curvas similares en la subsección 3.2: 
seguimientos de determinadas unidades léxicas a lo largo del corpus diacrónico de El País. Muestran ejemplos del 
comportamiento de unidades que consideramos neologismos, tales como \textit{teléfono móvil, teléfono fijo} y \textit{cambio climático}, unidades cuya 
frecuencia de uso muestra un incremento acusado en la línea del tiempo.

\begin{equation}
 f(x) = x^k 
\label{Ideal}
\end{equation}

\begin{figure}
\centering
\includegraphics[width=0.7\textwidth]{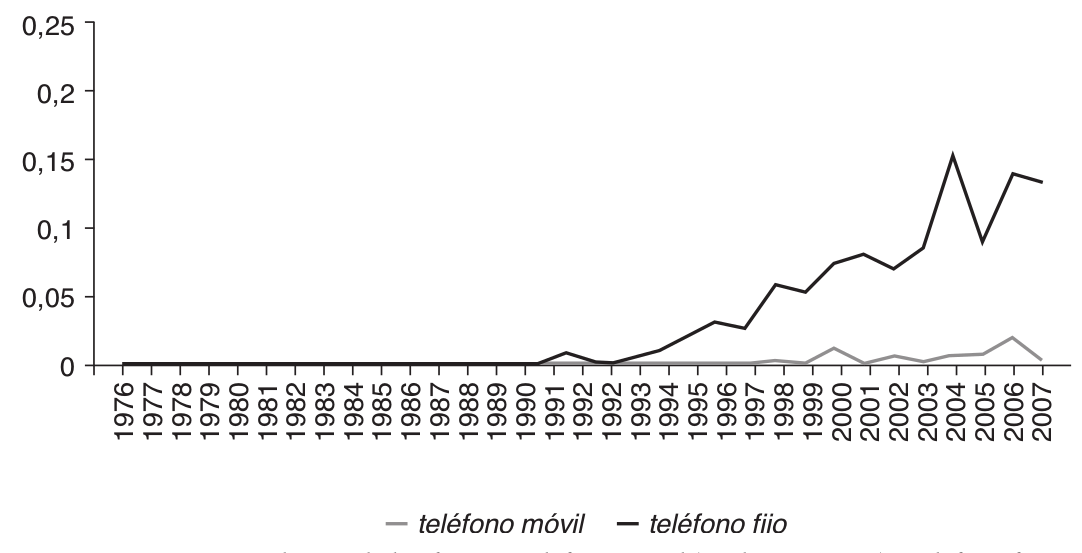}
\caption{\label{fig10}Distribución de las formas \textit{teléfono móvil} (curva superior) y \textit{teléfono fijo} (curva inferior) en el corpus diacrónico de El País.}
\end{figure}

\begin{figure}
\centering
\includegraphics[width=0.7\textwidth]{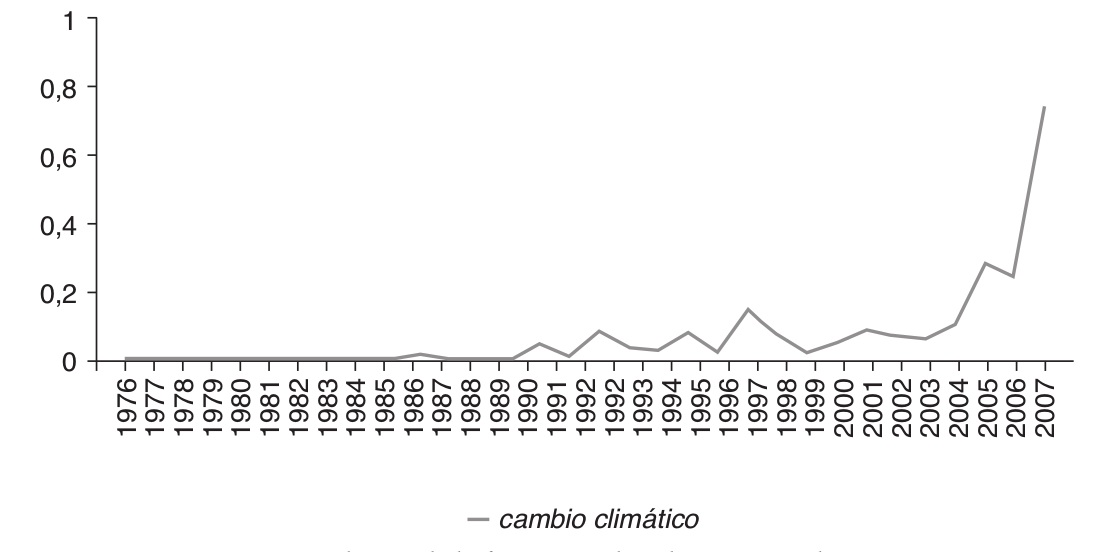}
\caption{\label{fig11}Distribución de la forma \textit{cambio climático} en el mismo corpus.}
\end{figure}

En dicho trabajo sobre extracción de neología definimos lo que sería la curva de comportamiento de un neologismo ideal o 
teórico, representada en la Figura 12 y definida en la Ecuación [12]. Se trata de una curva exponencial en el intervalo de años 
estudiado (en nuestros experimentos probamos con $k = 10$). El experimento consistió en tomar una muestra de $n$ unidades del corpus (Las unidades eran tanto palabras aisladas 
como secuencias de hasta cinco palabras de longitud) y ordenarlas de acuerdo con la distancia euclideana de sus curvas de 
frecuencia con la curva de este neologismo ideal. De este modo podemos obtener las unidades que se han ido incorporando a la 
lengua en los últimos años, unidades que luego se han de filtrar, ya que incluyen formas que no son neologismos, como es el 
caso de nombres propios o referentes que han adquirido notoriedad en los últimos años. 

Naturalmente, este sencillo método no 
resultaba eficaz en el caso de los neologismos semánticos, unidades que si bien son formalmente idénticas a otras formas de la 
lengua, se empiezan a utilizar con un significado diferente. Estas formas representan un desafío para la extracción automática 
con los métodos tradicionales, pero este mismo escenario es el que encontrábamos en la subsección 4.1.2, en la que 
clasificábamos contextos de aparición de unidades polisémicas. Es el caso, por ejemplo, de la forma \textit{palabra de honor}, que 
si bien tiene un uso literal, en el sentido de `hacer una promesa verbal', en los últimos años es cada vez más frecuente utilizarla 
para designar un determinado tipo de escote. Si bien su condición de neologismo para a este segundo sentido es discutible, ya 
que este tipo de escote no es nuevo, sí es nueva la masificación de este uso del término, y, por tanto, el ejemplo sigue siendo útil.

\begin{figure}
\centering
\includegraphics[width=0.7\textwidth]{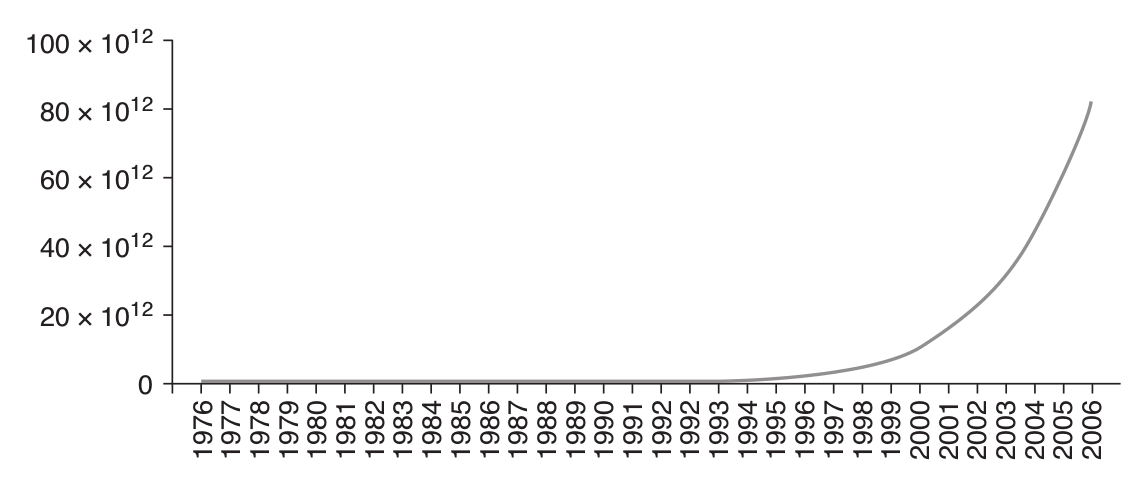}
\caption{\label{fig12}Gráfica del neologismo ideal.}
\end{figure}

Un algoritmo de clusterización o clustering similar al descrito en la subsección 3.2 es capaz de clasificar todos los contextos de 
aparición de la forma \textit{palabra de honor} en los archivos de El País y ofrecer dos clústeres con un nombre para cada uno. 
El clúster 1 se llama ``empeñar'' y el clúster 2 es llamado ``escotes'' . Cada uno de estos clústeres contiene una serie de 
unidades léxicas que 
conforman el entorno típico de las ocurrencias de la expresión en un sentido y en el otro. Así, en el clúster 1, tenemos unidades 
como: \textit{Astarloa, Barrionuevo, confederal, consentidas, credulidad, empeño, esclarece, Escudero, Fusté, Herrero, incite, 
inocencia, proclamó, quebrantamiento, reiterada}, etc. Estas formas se relacionan con el sentido literal, porque vemos que se trata de 
nombres propios de personajes públicos, para los que la credibilidad no debería ser irrelevante. En el caso del segundo clúster, 
en 
cambio, los vecinos típicos tienen relación con el mundo de la moda: \textit{cubres, drapeados, escotes, Gucci, marrón, modista, 
ojito, organiza, Swarovski, tonos}, etcétera.



\section{Conclusiones}

Se ha presentado aquí una visión amplia del cruce entre la lingüística y la estadística, con algunos ejemplos de técnicas que 
se pueden utilizar para el estudio del lenguaje. Estas técnicas se han acompañado, además, con ejemplos de aplicación 
concreta, como es el caso de la clasificación de documentos con o sin supervisión, así como la desambiguación de signos 
polisémicos y el descubrimiento de neología. Hubiera sido interesante mencionar otros ejemplos de aplicación práctica de estas 
técnicas, como la utilización de medidas de similitud para la comparación entre unidades léxicas de diferentes lenguas, es 
decir, la extracción de terminología bilingüe desde corpus no paralelos, o bien para la comparación de unidades léxicas de 
diferentes variedades dialectales. 

A priori, puede parecer que se trata de áreas de aplicación completamente diferentes, 
sobre todo para quien está acostumbrado a enfrentar tareas de este tipo con la incorporación de reglas explícitas que codifican 
conocimiento de la lengua o del dominio temático, así como información semántica extraída de diccionarios y ontologías, en el 
caso de la extracción de terminología, o corpus de exclusión lexicográficos, en el caso de la extracción de neología. La 
estadística, por el contrario, posibilita una manera diferente de concebir la lengua. Una investigación de la complejidad, 
pero desde una perspectiva integradora y simplificadora. Esto es así porque, desde el punto de vista estadístico, tareas y datos disímiles empiezan a 
parecer relacionados. A veces, los mismos métodos o las mismas formas de pensar se pueden aplicar a problemas que en 
principio parecían completamente diferentes. Concebimos, pues, la estadística como una «trans-disciplina». 

Para cerrar esta presentación, me gustaría remarcar que no hay que perder de vista el aspecto teórico. No estamos hablando sólo de «trucos 
ingenieriles» para resolver problemas prácticos que no tienen una relación intrínseca con la lingüística, como si estas 
soluciones estuvieran desprovistas de teoría. Está por ver si la estadística y la lingüística conforman disciplinas diferentes o si 
puede haber algo que llamamos una «sensibilidad estadística» en el análisis lingüístico, una manera de aproximarse a los datos, de advertir patrones, regularidades o tendencias en el cúmulo de los casos individuales donde el ojo humano no puede ver sino cantidad y diversidad.

\section*{Referencias}

{\small
\begin{itemize}[label={},leftmargin=*]
\item CABRÉ, M. T.; BACH, C.; DA CUNHA, I.; MORALES, A.; VIVALDI, J. (2009). Comparación de algunas características lingüísticas del discurso especializado frente al discurso general: el caso del discurso económico. XXVII Congreso de AESLA (Ciudad Real, 26-28 març 2009).

\item CASTORIADIS, C. (1975). La institución imaginaria de la sociedad. Buenos Aires: Tusquets.

\item CHURCH, K.; HANKS, P. (1990). «Word Association Norms, Mutual Information and Lexicography». Computational Linguistics, vol 16, núm. 1, p. 22-29.

\item DILTHEY, W. (1986). Introducción a las Ciencias del Espíritu. Madrid: Alianza.

\item EVERT, S. (2004). The Statistics ofWord Coocurrences. Tesis doctoral. Stuttgart: Universitat de Stuttgart. Institut fürMaschinelle Sprachverarbeitung, 2004.

\item HERDAN, G. (1964). Quantitative Linguistics.Washington: Butterworths.

\item MANDELBROT, B. (1961). «On the theory of word frequencies andMarkovianmodels of discourse». A: Structure of Language and itsMathematical Aspects. Symposia on Applied Mathematics. AmericanMathematical Society. Vol. 12, p. 190-219.

\item MANNING, C.; SCHÜTZE, H. (1999). Foundations of Statistical Natural Language Processing. MIT Press, 1999.

\item MOSTELLER, F.; WALLACE, D. (1984). Applied Bayesian and Classical Inference: the Case of the Federalist Papers. Nova York: Springer.

\item MULLER, C. (1973). Estadística Lingüística.Madrid: Gredos.

\item NAZAR, R. (2008). Diferencias cuantitativas entre referencia y sentido. Actas del XXVI
Congreso de AESLA. (Universitat d’Almeria, 3-5 de abril de 2008).

\item NAZAR, R. (2010). A Quantitative Approach to Concept Analysis. Tesis doctoral. Barcelona:
Universitat Pompeu Fabra. Institut Universitari de Lingüística Aplicada.

\item NAZAR, R; SÁNCHEZ POL,M. (2006). An Extremely Simple Authorship Attribution System. Second
European IAFL Conference on Forensic Linguistics / Language and the Law (Barcelona, 2006).

\item NAZAR, R.; VIDAL, V. (2008). Aproximación cuantitativa a la neología. I Congreso Internacional
de Neología en las lenguas románicas (Barcelona, 7-10 mayo 2008).

\item SEBASTIANI, F. (2000). Machine earning in automated text categorization. ACM Press,
vol. 34, núm. 1.

\item SHANNON, C. E. (1948). «Amathematical theory of communication». Bell SystemTechnical
Journal, vol. 27 (julio), p. 379-423.

\item SNOW, C. P. (1959 [1993]). The Two Cultures. Cambridge: Cambridge University Press.

\item TURELL,M. (2005). «Presentación». A: Lingüística forense, lengua y derecho: conceptos, métodos
y aplicaciones. Barcelona: Universitat Pompeu Fabra. Institut Universitari de
Ligüística Aplicada, p. 13-16.

\item VICKERS, B. (2002). Counterfeiting Shakespeare. Cambridge: Cambridge University Press.

\item VIVALDI, J. (2009). Corpus and exploitation tool: IULACT and bwanaNet. I Congreso Internacional
de Lingüística de Corpus (Murcia, 7-9 mayo 2009).

\end{itemize}
}

\end{document}